\documentclass[conference]{IEEEtran}
\IEEEoverridecommandlockouts

\usepackage{cite}

\usepackage{graphicx}

\usepackage[cmex10]{amsmath}

\usepackage{listings}
\usepackage{courier}
\usepackage{flushend}
\usepackage{color} %red, green, blue, yellow, cyan, magenta, black, white
\definecolor{mygreen}{RGB}{28,172,0} % color values Red, Green, Blue
\definecolor{mylilas}{RGB}{170,55,241}

\usepackage{courier}
\lstnewenvironment{code}[1][]%
{
              \noindent
              \minipage{\linewidth}
              \medskip
              \lstset{frame=lrtb,captionpos=b,basicstyle=\ttfamily\footnotesize,breaklines=true,frame=single,#1}}
{\endminipage}

\usepackage{array}

\usepackage[caption=false,font=footnotesize]{subfig}

\usepackage{multirow}
\usepackage{textcomp}

\makeatletter
\def\footnoterule{\relax%
   \kern-5pt
   \hbox to \columnwidth{\hfill\vrule width 0.5\columnwidth height 0.4pt\hfill}
   \kern4.6pt}
\makeatother

\author{Vijay Gadepally}

\begin{document}
\title{Hyperscaling Internet Graph Analysis with D4M on the MIT SuperCloud}

\author{\IEEEauthorblockN{Vijay Gadepally$^{\dagger}$, Jeremy Kepner$^{\dagger}$, Lauren
    Milechin$^{+}$, William Arcand$^{\dagger}$, David  Bestor$^{\dagger}$, Bill Bergeron$^{\dagger}$ 
\\ Chansup Byun$^{\dagger}$, Matthew Hubbell$^{\dagger}$, Micheal Houle$^{\dagger}$, Micheal Jones$^{\dagger}$, Peter Michaleas$^{\dagger}$, Julie Mullen$^{\dagger}$ 
\\ Andrew Prout$^{\dagger}$, Antonio Rosa$^{\dagger}$, Charles
Yee$^{\dagger}$, Siddharth Samsi$^{\dagger}$, Albert Reuther$^{\dagger}$}

\IEEEauthorblockA{$^\dagger$ MIT Lincoln Laboratory,   $^{+}$ MIT EAPS}
}

\maketitle

\let\thefootnote\relax\footnote{Vijay Gadepally is the corresponding
  author and can be reached at vijayg [at] ll.mit.edu. \\
     This material is based upon work supported by the Assistant
     Secretary of De- fense for Research and Engineering under Air
     Force Contract No. FA8721-05-C-0002 and/or FA8702-15-D-0001. Any
     opinions, findings, conclusions or recommendations expressed in
     this material are those of the author(s) and do not necessarily
     reflect the views of the Assistant Secretary of Defense for
     Research and Engineering.}

\begin{abstract}
Detecting anomalous behavior in network traffic is a major challenge due to the volume and velocity of network traffic. For example, a 10 Gigabit Ethernet connection can generate over 50 MB/s of packet headers. For global network providers, this challenge can be amplified by many orders of magnitude. Development of novel computer network traffic analytics requires: high level programming environments, massive amount of packet capture (PCAP) data, and diverse data products for ``at scale'' algorithm pipeline development.  D4M (Dynamic Distributed Dimensional Data Model) combines the power of sparse linear algebra, associative arrays, parallel processing, and distributed databases (such as SciDB and Apache Accumulo) to provide a scalable data and computation system that addresses the big data problems associated with network analytics development.  Combining D4M with the MIT SuperCloud manycore processors and parallel storage system enables network analysts to interactively process massive amounts of data in minutes.  To demonstrate these capabilities, we have implemented a representative analytics pipeline in D4M and benchmarked it on 96 hours of Gigabit PCAP data with MIT SuperCloud.  The entire pipeline from uncompressing the raw files to database ingest was implemented in 135 lines of D4M code and achieved speedups of over 20,000.
\end{abstract}

\IEEEpeerreviewmaketitle

\section{Introduction}
\label{sec:intro}

The rapid rise of sophisticated cyber threats is well documented and a growing threat to our information systems~\cite{kshetri2009positive,hale2002cybercrime}. Understanding internet phenomenology is challenging due to the variety of new threats and volume of new data being generated. To demonstrate the variety challenges, there are approximately 250,000 new malware programs registered each day~\cite{malware}, and a majority of all web traffic comes from bots, many of which are malicious in nature~\cite{bottraffic}.  These evolving vulnerabilities can lead to public safety concerns~\cite{lewis2002assessing} and economic impact: it is estimated that cyber attacks will cost nearly \$2 trillion in 2019~\cite{cybercost}. 

Enhancing this data variety challenge is the massive scale of internet and cyber networks. These ubiquitous networks form the basis of worldwide communication and it is estimated that in 2018, there will be almost 37 Terabytes per second  (TB/s) of Internet Protocol (IP) traffic~\cite{networking2016cisco}. To underscore the volume challenge in a more specific instance, consider the task of collecting and analyzing network traffic data to detect and remove botnets, networks of malicious computers that are controlled as a group. Network traffic data is typically collected using a packet capture appliance that intercepts packets moving through a network. These packets can be analyzed to look for broad trends across a network using packet metadata such as IP addresses, protocol, packet lengths, etc. that are stored in the packet header. As described in~\cite{feily2009survey,zeidanloo2010taxonomy}, there are a number of approaches to detecting botnets from network flow data such as anomaly detection algorithms that look for particular patterns in the movement of IP packets. For example, a large number of packets with source IP addresses communicating with particular destination IP addresses in a non-human pattern may indicate a botnet or command-and-control server. For large networks, capturing, storing and processing packet capture data can be a large challenge. Consider a 10 Gigabit Ethernet (10 GbE) link. Such a link can often have over 100,000 packets/second. With a header length of 40 bytes, this translates to nearly 50 MB/s of storage and processing of header information alone! Looking for patterns within such data rates can be quite a challenge. Of course, enterprises and internet service providers often have much larger network links and sustaining such rates is only possible with massive computing systems serviced by parallel filesystems and/or databases.

Development of novel computer network traffic analytics requires: high level programming environments, massive amount of packet capture (PCAP) data, and diverse data products for ``at scale'' algorithm pipeline development. In this article, we present our approach to developing a scalable internet analytics platform applied to a IP network data using the D4M (Dynamic Distributed Dimensional Data Model)~\cite{d4m2012} analytics environment and MIT SuperCloud~\cite{reuther2013llsupercloud} interactive computing environment. D4M combines the power of sparse linear algebra, associative arrays, parallel processing, and distributed databases (such as SciDB and Apache Accumulo) to provide a scalable data and computation system that addresses the big data problems associated with network analytics development. The MIT SuperCloud allows users to interactively process massive amounts of data in minutes using the software and environments most familiar to them. 

\begin{figure}[t]
\centering
\includegraphics[width=\linewidth]{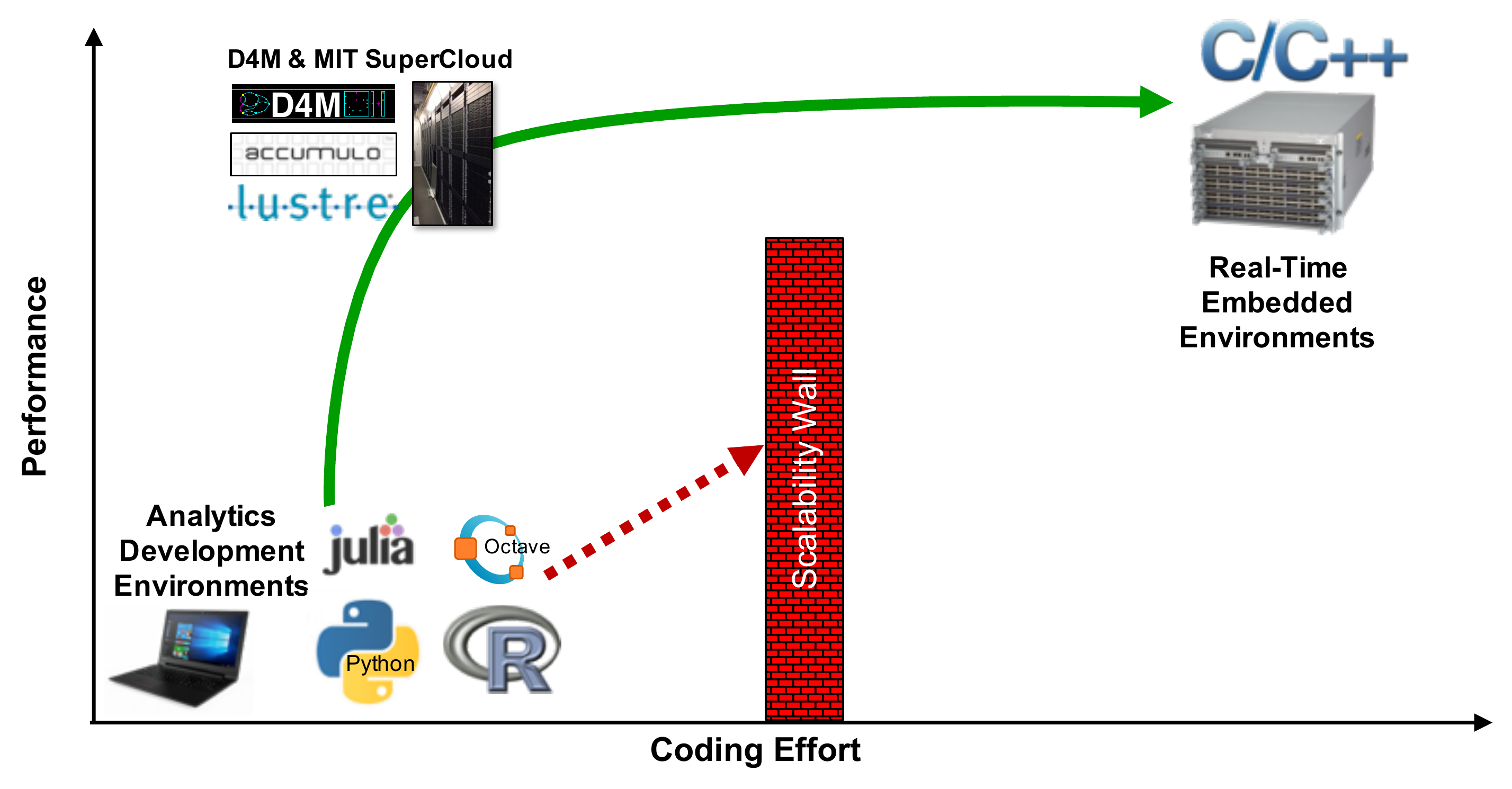}
\caption{D4M and MIT SuperCloud allows high performance without significantly compromising coding effort}\label{scalabilitywall}
\end{figure}

As depicted in Figure~\ref{scalabilitywall}, large enterprises typically develop analytics in environments such as Julia~\cite{bezanson2017julia} or machine learning environments like TensorFlow~\cite{abadi2016tensorflow} or Caffe~\cite{jia2014caffe}. While these have low coding effort and are useful for rapid analytic programming, they often come at the cost of lower relative performance. Deploying these algorithms to work at the scale of network traffic often implies translating these complex analytics into high performance languages such as C or C++ which requires significant coding effort. The time taken for this transition can be prohibitive for applications such as cyber analytics where there is a need for rapid deployment of analytics. Using MIT SuperCloud and D4M together allows cyber network analysts to overcome the scalability wall shown in Figure~\ref{scalabilitywall}. 

The rest of the article is organized as follows: Section~\ref{background} describes the tools used to develop a scalable network analysis pipeline, Section~\ref{implementation} details the implementation and Section~\ref{results} describes scaling and analytic results. Finally, we conclude in Section~\ref{conclusions}.

\section{Tools}
\label{background}

The scalable architecture described in this article leverages prior
work on D4M, Associative Arrays and the MIT SuperCloud computing platform.

\subsection{D4M}

The Dynamic Distributed Dimensional Data Model (D4M) is a software
library developed at MIT Lincoln Laboratory that is used in a number
of applications for processing large amounts of data. D4M is made up
of three components:

\begin{enumerate}
\item Support for a mathematical data object called associative arrays;
\item A schema that is used to represent unstructured data as associative arrays; and
\item A library of software tools for connecting associative arrays with database management systems ~\cite{gadepally2015d4m} such as Apache Accumulo, SciDB, mySQL, PostGRES.
\end{enumerate}

The D4M library is currently written to work in the analytic
environments of MATLAB, GNU Octave,  Julia~\cite{chen2016julia} and is currently being
implemented as a Python toolbox. To connect to database engines, D4M
can leverage high speed connectors or leverage existing
connectors.  D4M is a natural environment to match the scaling requirements of
network packet capture data and analytics.

\subsection{Associative Arrays}

Associative Arrays generalize matrices to better match the intuitions
of spreadsheets, databases, and tables, all while supporting the power
of linear algebra. The indices of an associative array can range over
arbitrary (though usually totally ordered and finite) sets, while the
entries of an associative array may lie in an arbitrary semiring,
which support addition and multiplication operations subject to most
of the familiar field laws (associativity, commutativity,
distributivity, identity, etc.).

This allows associative arrays to support many of the algebraic
features of matrices, including element-wise addition and
multiplication, as well as array multiplication and Kronecker
products.  By having its rows and columns meaningfully labeled,
associative arrays also allow these operations to be well-defined
between arrays of varying dimensions, unlike their matrix cousins.

\begin{figure*}[htbp]
\centering
\includegraphics[width=0.85\textwidth]{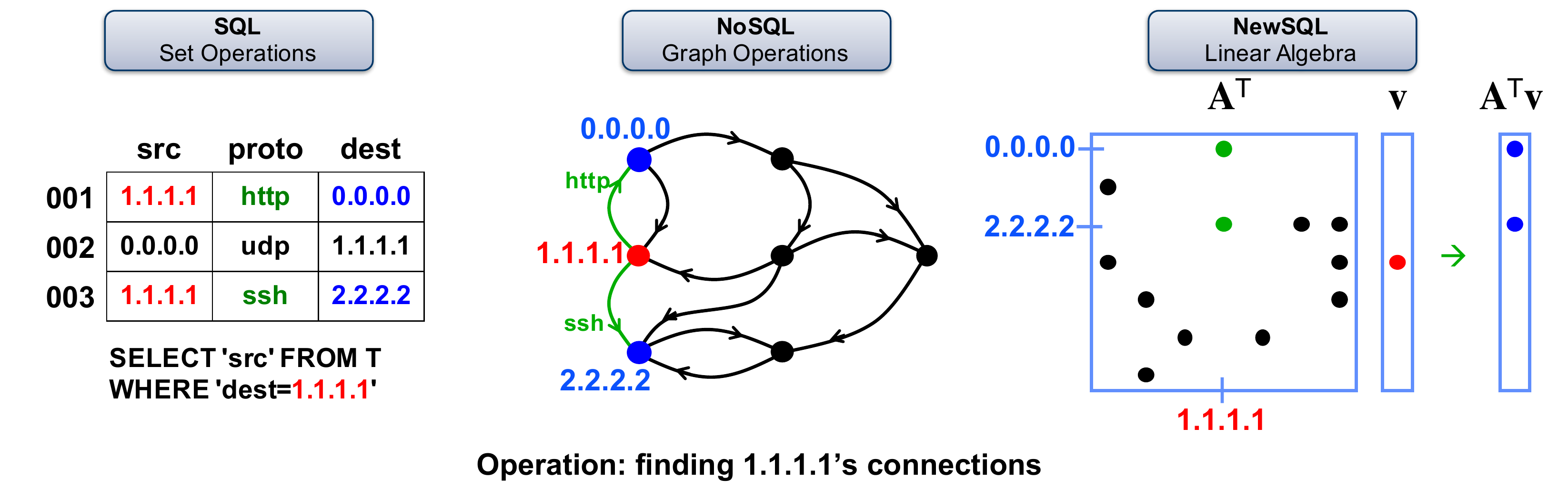}
\caption{Different data management and computing systems naturally
  support different types of operations. In this graphic, all of the
  systems are performing the same operation - looking for connections
  to 1.1.1.1 - in the language native to their system. Associative
  arrays provide a uniform mathematical framework capable of
  representing these operations.}\label{assocarray}
\end{figure*}

Associative array algebra can provide a uniform mathematical framework
to describe operations in SQL, noSQL and NewSQL databases\cite{kepner2016associative}. For
example, in Figure~\ref{assocarray}, we describe how one would express
the same operation in three different styles of database systems.

\subsection{MIT SuperCloud}

\begin{figure}[t]
\centering
\includegraphics[width=\linewidth]{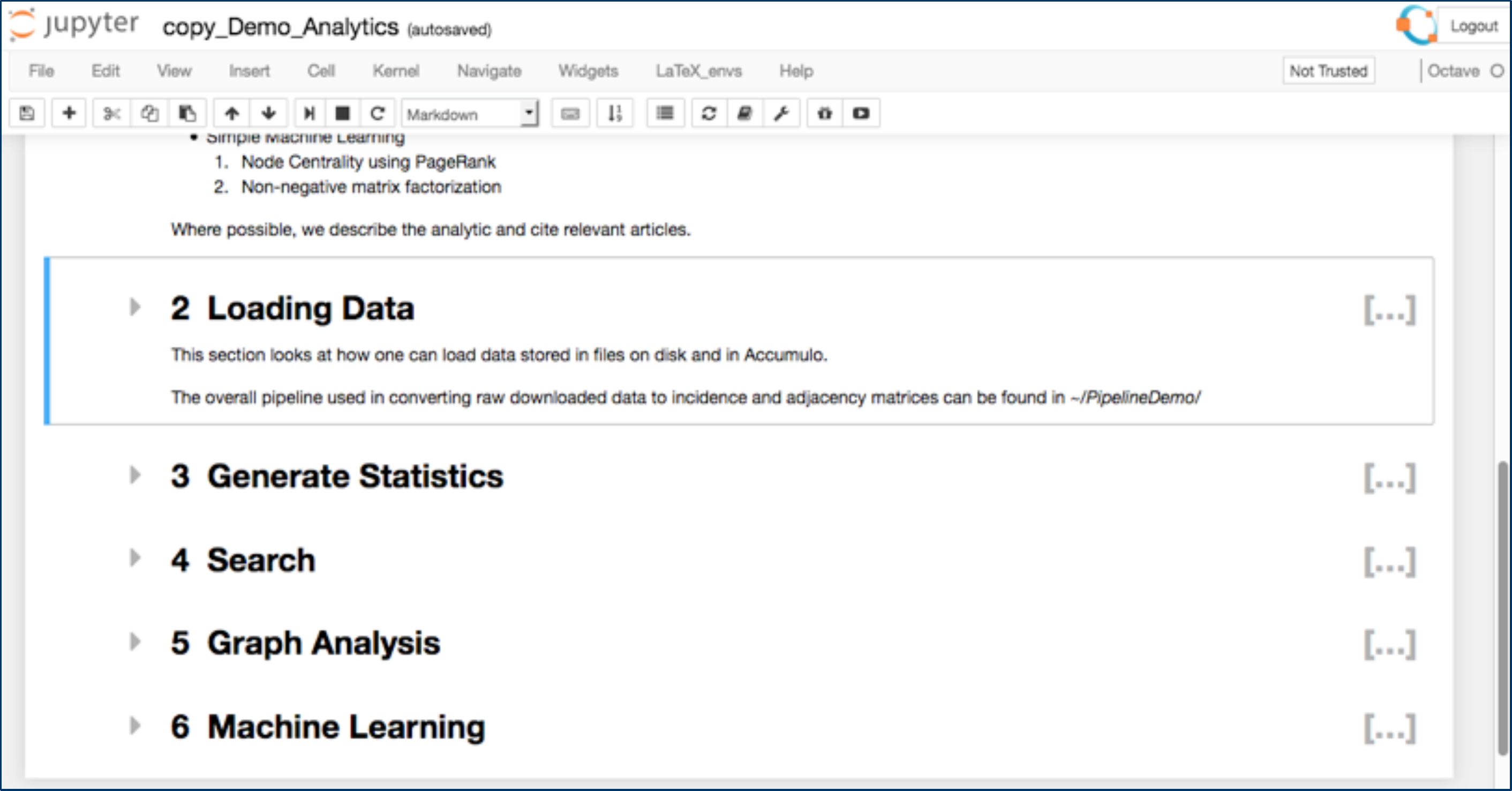}
\caption{Jupyter Notebook IDE available on the MIT SuperCloud.}\label{jupyter_ide}
\end{figure}

The MIT SuperCloud~\cite{reuther2013llsupercloud} is a high performance computing environment
developed at the Massachusetts Institute of Technology. Unlike
traditional supercomputing systems that are tuned for large-scale batch
processing, the MIT SuperCloud is designed for data scientists
interested in iterative analysis of machine learning and AI
workloads. Specific technologies such as interactive
databases~\cite{prout2015enabling}, and high performance analytic
IDEs such as
Jupyter~\cite{prout2017supercloud} provide a familiar environment for analysts. Figure~\ref{jupyter_ide} shows
the IDE used by MIT SuperCloud users.

Scaling jobs to tens of thousands of cores on the SuperCloud is straightforward due to support
for distributed computing paradigms such as
Map-Reduce~\cite{byun2016llmapreduce} and analytics development
environments tuned for high performance distributed computing~\cite{samsi2010matlab}.

\section{Developing Scalable Pipeline and Analytics}
\label{implementation}

Developing network analytics involves developing a pipeline that can
scale with the massive amount of data collected by packet capture
devices. This section details the network data used and computational pipeline developed.

\subsection{IP Network Traces}
\label{dataset}

IP network packets form the basic unit in which information is
transmitted across the internet. An individual network packet 
consists of a header and payload. The header consists of typical
information that one needs to correctly route a particular
packet such as source IP, destination IP, etc. Headers are typically
40 bytes in size. The second part of the packet is the payload or user
data. This consists of the actual data payload of the
packet. Payload information often consists of encrypted or sensitive
user data and most network analytics focus on packet headers for their
analysis. Listing~\ref{packetinfo} shows an example of the information (in
associative array form) encapsulated by an individual packet's header.

\begin{lstlisting}[label={packetinfo}]
(PacketID,frame.time_relative|0.000000000)     1
(PacketID,frame.time|2017 Apr 12 07:49:36.18828 EDT)     1
(PacketID,ip.dst|63.237.205.194)     1
(PacketID,ip.len|1500)     1
(PacketID,ip.proto|6)     1
(PacketID,ip.src|133.40.77.44)     1
(PacketID,tcp.dstport|55428)     1
(PacketID,tcp.flags|0x00000010)     1
(PacketID,tcp.srcport|80)     1
\end{lstlisting}

Using the packet header metadata, it is possible to determine clusters
of similar network flows~\cite{zhang2014analytics}, important activity using
centrality measures~\cite{franccois2011bottrack}, and anomalous
behavior based on clustering techniques~\cite{liu2013network}. We have
also applied domain agnostic techniques such as dimensional
analysis~\cite{gadepally2014big} and background modeling techniques for
power-law data~\cite{gadepally2015using}.

The MAWI Working Group (http://mawi.wide.ad.jp/mawi/) collects
and shares a variety of network trace data collected on the WIDE
network (http://www.wide.ad.jp/) backbone in
Japan~\cite{kato1999internet,sony2000traffic,fontugne2010mawilab}. The
working group has made a rich repository of data available for researchers interested in analyzing trends
in network traffic. The MAWI dataset provides a realistic view into
data collected by global internet
service providers.

\subsection{Computing Pipeline}
\label{pipeline}
\begin{figure*}[htbp]
\centering
\includegraphics[width=0.9\textwidth]{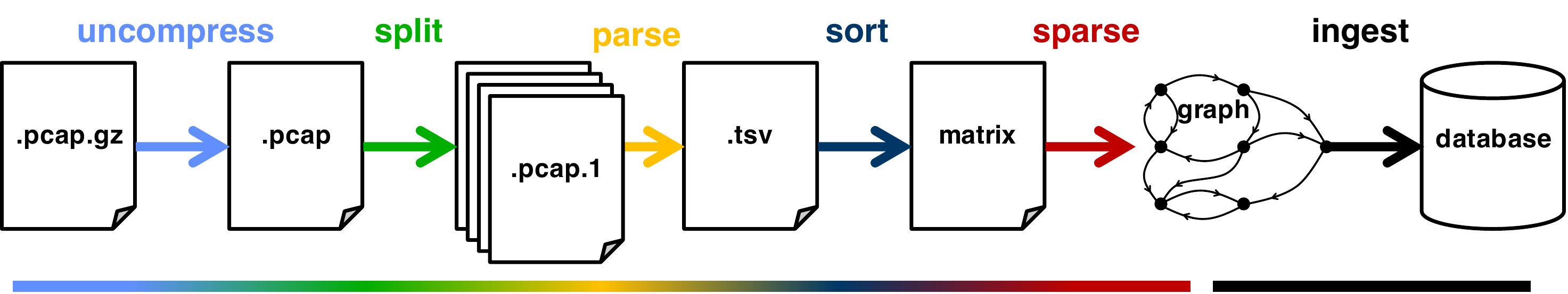}
\caption{Analytics pipeline used for processing network packet capture data.}\label{packetcapturepipeline}
\end{figure*}

Figure~\ref{packetcapturepipeline} describes the pipeline used to extract, store and
process packet capture data described in
Section~\ref{dataset}.

.

Each step of this pipeline is described below:

\begin{enumerate}

\item \textbf{Uncompress}: Data from packet capture appliances is
  often written in a binary compressed format. In this step, we uncompress each of the
  binary packet capture files in parallel in order to make it readable
  for subsequent processing.
\item \textbf{Split}: In order to make processing large data files
  efficient on high performance computing systems, each uncompressed binary file is read by a packet
  analyzer such as \texttt{tcpdump}~\cite{jacobson1989tcpdump} and split into smaller files. This
  step ensures that further processing is amenable to parallelization.
\item \textbf{Parse}: In this step, each of the split files is run
  through a network analyzer such as
  \texttt{tshark}~\cite{combs2012tshark} in order to convert the
  packet capture (pcap) data into to a human
  readable format. For our implementation, each output file is stored
  in a tab-separated value (TSV) format.
\item \textbf{Sort}: Each TSV packet capture file is converted into a
  dense associative array that is similar to a dense table in a relational
  database.
\item \textbf{Sparse}: Each dense array generated in the previous step
  is converted in an equivalent graph using the D4M schema.
\item \textbf{Ingest}: Each subgraph is inserted into a suitable
  database. For the purpose of our evaluation, we use the
  high-performance Apache Accumulo database~\cite{ingest2014}.

\end{enumerate}

In the following section, we describe specific settings and
performance details of the above pipeline.

\section{Experimental Results}
\label{results}

In order to test the pipeline of Figure~\ref{packetcapturepipeline},
we use data products made available by the MAWI working group. The dataset we use called a ``Day in the
Life'' (DITL) internet traces, consists of 4 days (96 hours) of 1 Gigabit packet capture
headers collected on two days in 2015 and two days in 2017. 

In total, the raw data in compressed form, is approximately 700 GB. When converted to an
analyst-friendly form by uncompressing, parsing, and sorting, the data expands to approximately 20
TB. Scaling performance analysis was performed on MIT Lincoln
Laboratory Supercomputing Center's system. This system consists of 
650 nodes with Intel Xeon 64-core processors and 180 nodes with 32-core AMD
Opteron processors. For our scaling experiments, processing is performed on
Xeon-64 nodes and databases are operated on AMD Opteron nodes. For the
results presented below, the maximum processing size was performed on
385 Intel Xeon-64 nodes (24,640 cores) and the largest database instance was
distributed across 128 AMD
Opteron nodes.

In the subsections below, we described the implementation details of
the pipeline presented in Section~\ref{pipeline}.

\subsection{Step1: Uncompress Raw Data}

In this step, we take 385 compressed input files (corresponding to the
number of computing nodes used in the experiment) and convert them to
385 uncompressed output files. Each input and output file corresponds
with roughly 15 minutes of network flows. Each 2GB file expands
to 6GB after uncompressing which translates to an increase from 700GB
to approximately 2.3 TB. The maximum speedup is largely impacted by the
number of input files and eventually limited by the file system I/O
speed. 

The code snippet below shows the D4M code used to uncompress a single
\textit{.pcap} file. In this snippet, \texttt{dataDir} and \texttt{dataDOM} corresponds with file
locations on the system and iFile corresponds to the compressed file to
be processed by each node:

\begin{lstlisting}
uncompressCommand = ['gunzip -k ' dataDIR dataDOM '/' iFile '.pcap.gz']
system(uncompressCommand);
\end{lstlisting}

\subsection{Step 2: Split Uncompressed Files}

Once the uncompressed output files are generated from the previous step,
we split these files into smaller chunks in order to optimize later steps in
the pipeline. We first use \texttt{tcpdump} to convert the 385 binary
\textit{.pcap} files into ASCII versions, then split these files into
into approximately 500,000 smaller output \textit{.pcap} files
appended with a split ID. Similar to the previous step, the maximum
speedup is largely impacted by the number of input files (which should
closely match the number of processing nodes) and
eventually limited by file system I/O.

The code snippet below shows the code used to convert from
binary to ASCII and split the input \textit{.pcap} files. The \texttt{splitSize} was set to
be 5 MB.

\begin{lstlisting}
splitCommand = ['/usr/sbin/tcpdump -C ' splitSize ' -r ' dataDIR dataDOM '/’
         iFile '.pcap -w ' dataDIR dataDOM '/' iFile '/' iFileName '.pcap. 2>&1'];
[status,result] = system(splitCommand);
\end{lstlisting}

\subsection{Step 3: Parse Split Files}

With the approximately 500,000 split and uncompressed \textit{.pcap} files, we convert
these files into a human readable format using a tool such as
\texttt{tshark}~\cite{combs2012tshark}. Using \texttt{tshark}, we convert these
\textit{.pcap} into a tab separated value (TSV) format while also
filtering the headers for the fields shown in
Section~\ref{dataset}. Each output TSV file at this stage is
approximately 5MB in size (for a total of 2.3 TB across all files) and
each TSV file corresponds to rougly 1 second of network flow data. The maximum speedup of this step is
limited by the number of cores available for parsing.

The code snippet below describes the D4M operations used for parsing
the split files:

\begin{lstlisting}
savetsvCommand = ['/usr/sbin/tshark -r ' dataDIR dataDOM '/' iFile '/' ijFile ...
          ' -n -Tfields ' [' -e ' strrep(tsvHeader(1:end-1),tab,' -e ')] ...
          ' > '  parseDIR dataDOM '/' iFile '/' ijFile '.tsv'];
system(savetsvCommand);
\end{lstlisting}

\subsection{Step 4: Dense Array Construction (Sort)}

In order to convert the 500,000 files in the previous step to a format
amenable for further processing, we use D4M to convert these TSV files
into associative array format (which also sorts the data
during construction). Each of the 5 MB input files expands to roughly
50 MB (total of 20 TB) during this step. At this point, data is human
readable and ready to construct the network graph. At this stage, the
maximum speedup is limited by the number of cores.

The code snippet below describes the D4M syntax to load the input
files, restructure the time field, construct the associative array and
save the resultant sorted array to disk.

\begin{lstlisting}
ijTSVstr = StrFileRead([parseDIR dataDOM '/' iFile '/' ijFile '.tsv']);
ijTSVstr = [tsvHeader  ijTSVstr];
A = CSVstr2assoc(ijTSVstr,nl,tab);
[r c v] = A(:,'frame.time,');         % Restructure time field.            
tMatStr = Str2mat(v);
At = Assoc(r,'frame.time,',Mat2str(tMatStr(:,[10:14 2:7 14:36 38])));
A = (A - At) + At;
A = putRow(A,CatStr(Row(A),'.',[ijFile '.A.mat,']));
save([parseDIR dataDOM '/' iFile '/' ijFile '.tsv.A.mat'],'A');
\end{lstlisting}

\subsection{Step 5: Graph Construction (Sparse)}

With the dense associative arrays from the previous step stored on
disk, we can now, in parallel, generate the sparse version of the
network graph. This sparse representation directly corresponds to the
incidence matrix of the graph. Each of the 50 MB input associative
array is converted to a sparse representation using the D4M
schema. The resultant output file is saved to disk for database
insertion. As in the previous step, the maximum speedup of this step
is limited by the number of cores.

The code snippet below describes the D4M syntax for loading in the
output array from Step 4 and converting it to a sparse representation:

\begin{lstlisting}
load([parseDIR dataDOM '/' iFile '/' ijFile '.tsv.A.mat'],'A');
E = val2col(A,'|');
save([parseDIR dataDOM '/' iFile '/' ijFile '.tsv.A.mat.E.mat'],'E');
\end{lstlisting}

\subsection{Step 6: Ingest}

With the sparse data products of the previous step, it is easy to use
D4M to directly insert this data into Apache Accumulo. Our prior work
has demonstrated that Accumulo is capable of extremely high ingest
rates suitable for applications such as internet traffic analysis. In
our experiment, we create various Accumulo instances with different
configurations in order to test scalability. For our testing, we
deploy Accumulo on 32-core AMD Opteron nodes. To test scalability, we
use 1, 4, 16 node instances. For larger database instances, we found
that running multiple database systems over 16 nodes was more
efficient than larger Accumulo instance (i.e., 2, 4, 8 databases
running in parallel each with 16 nodes rather than 32, 64, and 128
node instances).  

Each of the 385 Xeon-64 nodes is
responsible for loading a subset of the 500,000 sparse arrays from the
previous step into an Accumulo instance (or a particular Accumulo
instance in the case where we have multiple databases in parallel). The maximum speedup for this step is limited by the
number of Accumulo cores available.

The code snippet below describes the D4M syntax for loading the
incidence matrix file, inserting into a table called Tedge,
generating the degree table and inserting it into TedgeDeg. Details
about the general schema and table design can be found in~\cite{kepner2013d4m2schema}.

\begin{lstlisting}
load([parseDIR dataDOM '/' iFile '/' ijFile '.tsv.A.mat.E.mat'],'E');
put(Tedge,putVal(E,'1,'));
Edeg = putCol(sum(E.',2),'degree,');
put(TedgeDeg,num2str(Edeg));
\end{lstlisting}

\subsection{Performance Analysis}

To assess the performance of each step, the D4M code included
timers. For each of the first five steps of the pipeline, the time
measured includes the time for reading the file from disk, performing
the operation and writing the file back to disk. For the insertion
step, we measure the time taken to load the file and insert into
Accumulo. For each of the experiments, we fix the data size and
compute the speedup relative to the time taken for a single core to
perform the task.

\begin{figure}[t]
\centering
\includegraphics[width=\linewidth]{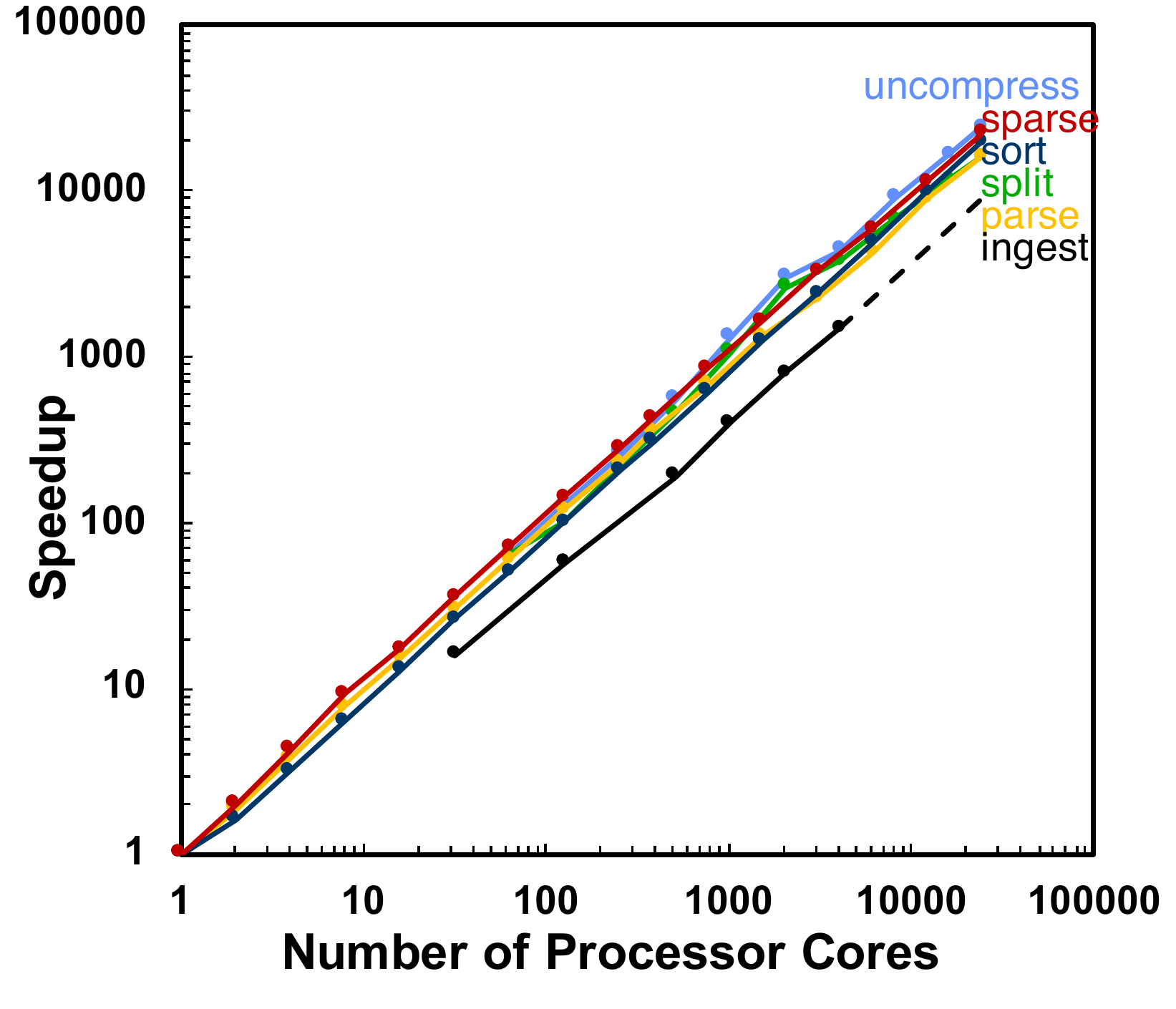}
\caption{Scaling Performance using D4M and MIT SuperCloud. The x-axis
  indicate the number of processing cores and the y-axis the
  resulting speedup.}\label{insertPerformance}
\end{figure}

Figure~\ref{insertPerformance} describes the speedup associated
with a varying number of processing cores for each of the steps of the
processing pipeline. Each line in the figure is color coded according
to the color of the step in Figure~\ref{packetcapturepipeline}. The
ingest line starts at 32 database cores and ends at 4096 database
cores (corresponding to 8 x 16-node Accumulo databases).

As seen in Figure~\ref{insertPerformance}, for Steps 1-5 of the
pipeline, increasing the number of cores leads to a near linear
speedup. For the database ingest, the speedup is largely limited by
the number of Accumulo cores available (4096 is the maximum number of
Accumulo cores in our experiment). The entire pipeline from uncompressing the raw data to database ingest
was implemented in approximately 135 lines of D4M code.

\section{Conclusions}
\label{conclusions}

Network and cyber security of the future will largely rely on massive quantities of data. Internet network analysis will continue to be challenged by the fast pace of analytic changes coupled with massive quantities of data. In order to address these challenges, it is important that researchers leverage high level programming environments that simplify analytic development along with computing platforms that support high-performance analysis.  In this article, we describe our approach to developing such a toolbox based on D4M and MIT SuperCloud. We describe our approach to using this infrastructure to develop a processing pipeline for IP traces collected by the MAWI working group. As is described in the article, our system allows researchers to develop scalable processing pipelines without compromising coding effort.

\section*{Acknowledgment}

The authors acknowledge the following individuals for their
help in understanding the MAWI dataset: Koichi Suzuki, Kenji Takahashi, Michitoshi Yoshida, Bo Hu and Shohei
Araki. The authors also wish to acknowledge the support of MIT SuperCloud team
and Hayden Jananthan.



%\bibliographystyle{IEEEtran}
%\IEEEtriggeratref{6}
%\balance
\bibliography{cyberd4m}
\bibliographystyle{IEEEtran}

\end{document}